\documentclass[prd,twocolumn, showpacs,amsmath,amssymb,reprint,preprintnumbers,nofootinbib]{revtex4-1}

\RequirePackage[colorlinks=true
,urlcolor=blue
,anchorcolor=blue
,citecolor=blue
,filecolor=blue
,linkcolor=blue
,menucolor=blue
,pagecolor=blue
,linktocpage=true
,pdfproducer=medialab
,pdfa=true
]{hyperref}

\usepackage{amsmath,amssymb,amsthm,amsfonts}
\usepackage{graphicx,tabularx}
\usepackage{color}
\usepackage{multirow}  

\newcommand{\be}{\begin{equation} \begin{aligned}}
\newcommand{\ee}{\end{aligned} \end{equation}}

\begin{document}
\title{Towards an Analytic Construction of the Wavefunction of Boson Stars}

\preprint{UCI-HEP-TR-2017-04}

\author{Felix Kling}
\email{fkling@uci.edu}
\affiliation{Department of Physics and Astronomy, University of
California, Irvine, CA 92697, USA}

\author{Arvind Rajaraman}
\email{arajaram@uci.edu}
\affiliation{Department of Physics and Astronomy, University of
California, Irvine, CA 92697, USA}

\begin{abstract}
Light scalar fields can form gravitationally bound compact objects called boson stars. The profile of boson stars  in the Newtonian limit is described by the Gross-Pitaevskii-Poisson equations. We present a semi-analytic solution to these equations and construct the profile of boson stars formed by a non-interacting scalar field. Our solution is stable with respect to numerical errors and has accuracy better than $10^{-6}$ over the entire  range.  
\end{abstract}

\maketitle

\section{Introduction}

Axions are an attractive solution 
\cite{1977PhRvL..38.1440P, 1977PhRvD..16.1791P, 1978PhRvL..40..223W, 1978PhRvL..40..279W, 
1981PhLB..104..199D}to the strong CP problem of QCD \cite{1976PhLB...63..334C, 1976PhRvL..37..172J}, and also provide an attractive and natural dark matter candidate (for reviews see
\cite{Kim:2008hd, Cheng:1987gp}). This has motivated multiple search strategies for axions and axion-like particles \cite{1983PhRvL..51.1415S, 1985PhRvD..32.2988S, 2010PhRvL.104d1301A, 2011PhRvD..84l1302H, Rybka:2014cya, Sikivie:2013laa, 2012PhRvD..85c5018B, Horns:2012jf, Budker:2013hfa, Graham:2013gfa, 1999NIMPA.425..480Z,Stadnik:2013raa}. It has also been argued that ultra-light axions \cite{Hui:2016ltb} can solve problems encountered by the usual CDM models \cite{2011AJ....141..193O, 1993MNRAS.264..201K, 1999ApJ...522...82K, 2000PhRvL..84.4525K}.

For all these purposes, it is important to understand the spatial structure of axion-like particles 
if they constitute dark matter. Most importantly, it is crucial to know whether or not these particles 
clump into compact objects (see \cite{Liebling:2012fv} for a review). The Jeans instability 
\cite{1985MNRAS.215..575K, Bianchi:1990mha, 1994PhRvL..72.2516S, Mielke:2000mh, 1986PhRvL..57.2485C} indicates that a uniform density of axions is unstable, indicating the formation of large compact objects (sometimes called boson stars \cite{1968PhRv..172.1331K, Ruffini:1969qy}). The boson stars are prevented from completely collapsing by a bosonic analogue of the Fermi pressure in white dwarfs \cite{1926MNRAS..87..114F}. 

Once the boson stars form, further cosmological evolution can occur by scattering of these boson stars 
off other boson stars, as well as baryonic matter \cite{Eby:2017xaw}. Such scatterings 
may either enhance or decrease the number of these objects. To analyze these scatterings, we 
must have a detailed understanding of the bound states, including their energies and profiles. 

At least for fairly dilute\footnote{There is a different set of solutions with large density: dense boson stars 
\cite{Braaten:2016kzc, Braaten:2015eeu,Eby:2016cnq,Bai:2016wpg}.} systems, the compact objects are bound states of a nonlinear Gross-Pitaevskii-Poisson equations, which we re-derive below. Numerical solutions to these equation have found the bound state energies and mass-radius relation, both in the cases with no self-interactions, as well as including self interactions, either attractive or repulsive 
\cite{Ruffini:1969qy, Barranco:2010ib, 2007JCAP...06..025B, 1989PhRvA..39.4207M, 1996PhRvD..53.2236L, 2000NewA....5..103G, 2003PhRvD..68b3511A,Chavanis:2011zm, Eby:2014fya,Eby:2015hsq,Yang:2015paa,Kan:2017uhj}. 
However, the profiles are only available numerically; they are computationally expensive to find and are difficult to extend to  perturbations of the boson stars. Furthermore, they tend to have numerical instabilities in the tails of the profile. The author of \cite{Chavanis:2011zi} follows a different approach: using a 
simple Gaussian ansatz for the density profile they were able to obtain the mass-radius relation and 
ground-state energy within a 10\% deviation of the numerical solution. The Gross-Pitaevskii-Poisson equations have also been studied in the context of quantum state reduction by  
\cite{Penrose1998, Moroz:1998dh, Tod1992}.

In this paper, we introduce a combination of analytical and numerical methods to find approximate 
solutions to the Gross-Pitaevskii-Poisson equations. We illustrate this for the case of no self-interactions 
among the bosons (the interacting case will be treated in an upcoming paper \cite{Kling:2017hjm}). We also perform a detailed analysis of the accuracy of our results. We show that our method is much less computationally expensive than previous approaches, nevertheless we demonstrate that we find excellent agreement with a full numerical solution over the entire parameter range. Furthermore, our method is numerically stable and does not diverge far away from the star, which can occur for a purely numerical approach.

\section{The Gross-Pitaevskii-Poisson Equations}

\subsection{Derivation}

Let us consider a complex scalar field $\phi(\vec{r},t)$ described by the Lagrangian
\be
\mathcal{L} = g^{\mu\nu} (\partial_\mu \phi^*)(\partial_\nu \phi) - m^2 (\phi^* \phi) - 
\frac{\lambda}{2} (\phi^* \phi)^2
\ee
We can then expand the field $\phi$ in spherical harmonics\footnote{Note that this expansion allows us to choose 
all $R_{nl}(r)$ to be real.}
\be
\phi (\vec{r},t) 
= \sum_{nlm} R_{nl}(r) Y_{lm}(\theta,\phi) e^{-i E_{nlm} t } .
\ee
We assume that only the ground state $(n,l,m)=(1,0,0)$ is populated. In this case the field 
takes the simple form $\phi (\vec{r},t) = (2E/N)^{-\frac{1}{2}}\psi(r) e^{-i E t}$, where we denote the ground-state energy as $E$. The real function $\psi(r)$ describes the radial profile of the star and is sometimes called the  wavefunction. We have chosen a normalization $\int \psi^2 dV = 1$, which allows us to identify $\psi^2$ with the probability density. 
Let us further make two simplifying assumptions
\begin{description}
\item [Newtonian Gravity] The field is only weakly coupled to gravity such that we can use a Newtonian 
approximation. This allows us to introduce the Newtonian potential $\Phi$ in the metric $g_{\mu\nu}= 
\text{diag}(1+2\Phi,-1,-1,-1)$. The Newtonian potential is related to the energy density via the 
Poisson equation $\nabla^2 \Phi = 4 \pi G \rho$.
 \item [Non-Relativistic] The ground state is non-relativistic. In this case we can 
write $E = m + e$ with binding energy $e \ll m$. This implies $e\psi, \Phi\psi, \nabla \psi \ll  m \psi $. 
\end{description}
\vspace*{0.1em}

The equation of motion is the 
Klein-Gordon equation, $\Box \phi + m^2 \phi+\lambda |\phi|^2 \phi =0$, which we can write 
in terms of the wavefunction 
as 
\be
(1+2\Phi)^{-1 }\partial_t^2 \psi - \nabla^2 \psi +m^2 \psi + \frac{N\lambda}{2m} \psi^3= 0 .
\label{realscalar-kleingordonequation}
\ee
In the non-relativistic approximation we can write
\be
e \psi = -\frac{1}{2m} \nabla^2 \psi + m \Phi \psi +   \frac{N \lambda}{4 m^2 } \psi^3
\label{realscalar-schrodingerequation}
\ee
and recover the Schr{\"o}dinger-type equation. For our non-relativistic approximation to be 
consistent, the last term should be sufficiently small i.e. $ \frac{N \lambda}{4 m^2 } \psi^2 \ll m$.

The energy density of the complex scalar field is
\be
\rho = |\partial_t \phi|^2 + |\nabla \phi|^2+ m^2 |\phi|^2 + \lambda |\phi|^4
\approx N m \psi^2
\ee
where we used the non-relativistic approximation in the last step. Newtons equation therefore 
takes the simple form
\be
\nabla^2 \Phi = 4 \pi G Nm \psi^2 
\label{realscalar-newtonequation}
\ee

\subsection{Scaling Symmetry of the Gross-Pitaevskii-Poisson System}

As derived above, the ground state of the self-gravitating boson star in the 
non-relativistic limit can be described by a wavefunction $\psi(r)$ and a gravitational potential 
$\Phi(r)$ satisfying the Gross-Pitaevskii-Poisson equations given in 
(\ref{realscalar-schrodingerequation}) and (\ref{realscalar-newtonequation}). For the remainder of 
this paper, we will focus on the non-interacting case $\lambda=0$, and postpone the discussion of 
finite self-interactions $\lambda\neq0$ to a separate study \cite{Kling:2017hjm}. 
For simplification, let us introduce the following dimensionless variables, 
\be
x=2rm,  \;\; V = \frac{e}{2m}-\frac{\Phi}{2} , \;\;\text{and}\;\; S = \sqrt{\frac{\pi G N}{2m}} \psi
\label{gpp-definevariables}
\ee
The equations (\ref{realscalar-schrodingerequation}) and (\ref{realscalar-newtonequation}) become
\be
\nabla^2 V = - S^2 \quad \text{and} \quad   \nabla^2 S =  - V S .
\label{gpp-dimensionless}
\ee
The wavefunction normalization condition $\int \psi^2 dV =1$ can be rewritten as  
\be
\int_0^\infty x^2 S^2 dx = G M m,
\label{gpp-mass}
\ee
where we introduced the star mass $M=Nm$. These equations describe the hydrostatic equilibrium 
of the boson star, in which the gravitational attraction caused by the potential $V$ is 
balanced by a repulsive quantum pressure. This quantum pressure arises from Heisenberg's uncertainty 
principle and prevents the system from gravitational collapse.

Let us note that Eq.~\ref{gpp-dimensionless} and \ref{gpp-mass} are invariant under the scaling 
\be
x \to \frac{x}{f}, \quad S \to f^2 S, \quad V \to f^2 V, \quad M \to f M.
\label{gpp-scaling}
\ee 
where $f$ is a scaling factor. This implies that solutions corresponding to different  
masses $M$ can be related to a unique solution by rescaling. To obtain the unique 
solution, we have to fix the scale by choosing a reference scale $k$. Although there are many 
different ways to fix $k$, a particularly useful choice for our discussion\footnote{For the 
numerical integration in appendix \ref{app-numerics}, we will choose a different reference scale 
$k$ which sets $V(0)=1$.} is to set $-k^2 = V(\infty)=\frac{e}{2m}$ which transforms as $k \to f k$. We 
can then introduce the scale invariant coordinate $z$, wavefunction $s$, potential $v$ and mass 
$\beta$ via
\be
z = kx, \;\; S = k^2 s, \;\; V = k^2 v, \;\; GMm=2k\beta.
\label{gpp-rescale}
\ee
Note that $S$ and $V$ are functions of the scaling dependent coordinate $x$ while $s$ and $v$ are 
functions of the scale independent coordinate $z$. Using the scale independent variables, we can 
write the Gross-Pitaevskii-Poisson equations as
\be
\nabla^2 s = - sv \;\;\; \text{and}\;\; \nabla^2 v = -s^2 \;\;\;\text{with}\;\; v(\infty)=-1.
\label{gpp-scaleinvariantgpp}
\ee  
To obtain the solution corresponding to a boson star with mass $M$, we then have to perform the 
rescaling given in Eq.~\ref{gpp-rescale} with $k =\frac{GMm}{2\beta}$. In the following section, we 
will obtain an approximate analytical form for $s,v$ and the mass parameter $\beta$. 

\section{Series Expansions}

It has been shown \cite{Tod1992,Choquard:2007aa} that the Gross-Pitaevskii-Poisson system given 
by (\ref{gpp-scaleinvariantgpp}) has a unique square normalizable solution for $s,v$ with $s>0$ 
for all values of $z$. The authors of \cite{Moroz:1998dh} also provide a numerical solution. However, 
the authors have also shown that this numerical solution will diverge at some finite value of $z$ and 
can therefore not be used to describe the profile at large radius. The author of 
\cite{Chavanis:2011zi} follows a different approach and approximates the density 
profile by a Gaussian, which is able to approximately reproduce thermodynamical properties of 
the boson star but otherwise fails to describe the profile, in particular at large radius. We 
attempt to solve this problem by providing an analytical expression for $s$ and $v$ which 
describes the profile at all radii with high precision. 

\subsection{Expansion at Small Radius: $z=0$}

Since both $s$ and $v$ are well behaved around $z=0$, we can expand them in a series expansion
\be
s = \sum_{n=0}^{\infty} s_n z^n \;\;\text{and}\;\;
v = \sum_{n=0}^{\infty} v_n z^n\ .
\label{smallz-expansion}
\ee
We can then write the Laplacian on the left hand side of Eq.~\ref{gpp-scaleinvariantgpp} as
\be
\nabla s = s'' + \frac{2}{z}s' 
= \sum_{n=0}^{\infty} (n+2)(n+3) \; s_{n+2}\;  z^{n}.
\ee
and the right hand side as a Cauchy product
\be
sv =  \Bigg[ \sum_{i=0}^\infty s_i z^i \Bigg] \Bigg[ \sum_{j=0}^\infty v_j z^j \Bigg]
=   \sum_{n=0}^\infty  \sum_{m=0}^{n} s_m v_{n-m} z^n.
\ee
By matching the coefficients in Eq. \ref{gpp-scaleinvariantgpp} we obtain the recursion relations
\be
(n+2)(n+3)  s_{n+2} &= -\sum_{m=0}^{n} s_m v_{n-m} ,\\
(n+2)(n+3)  v_{n+2} &= -\sum_{m=0}^{n} s_m s_{n-m} .
\ee
Note that requiring the profile to be smooth at the origin implies $s_1 = v_1 = 0$ and therefore 
also all odd coefficients $s_{2n+1}, v_{2n+1}$ vanish. The profile at small radius $z$ can 
therefore be fully parametrized in terms of the wavefunction and potential at the origin: $s_0$ 
and $v_0$. We can obtain $s_0$ and $v_0$ from a fit to the numerical solution, as discussed in 
appendix \ref{app-numerics}. 

\begin{figure}[t!]
\centering
\includegraphics[width=0.45\textwidth]{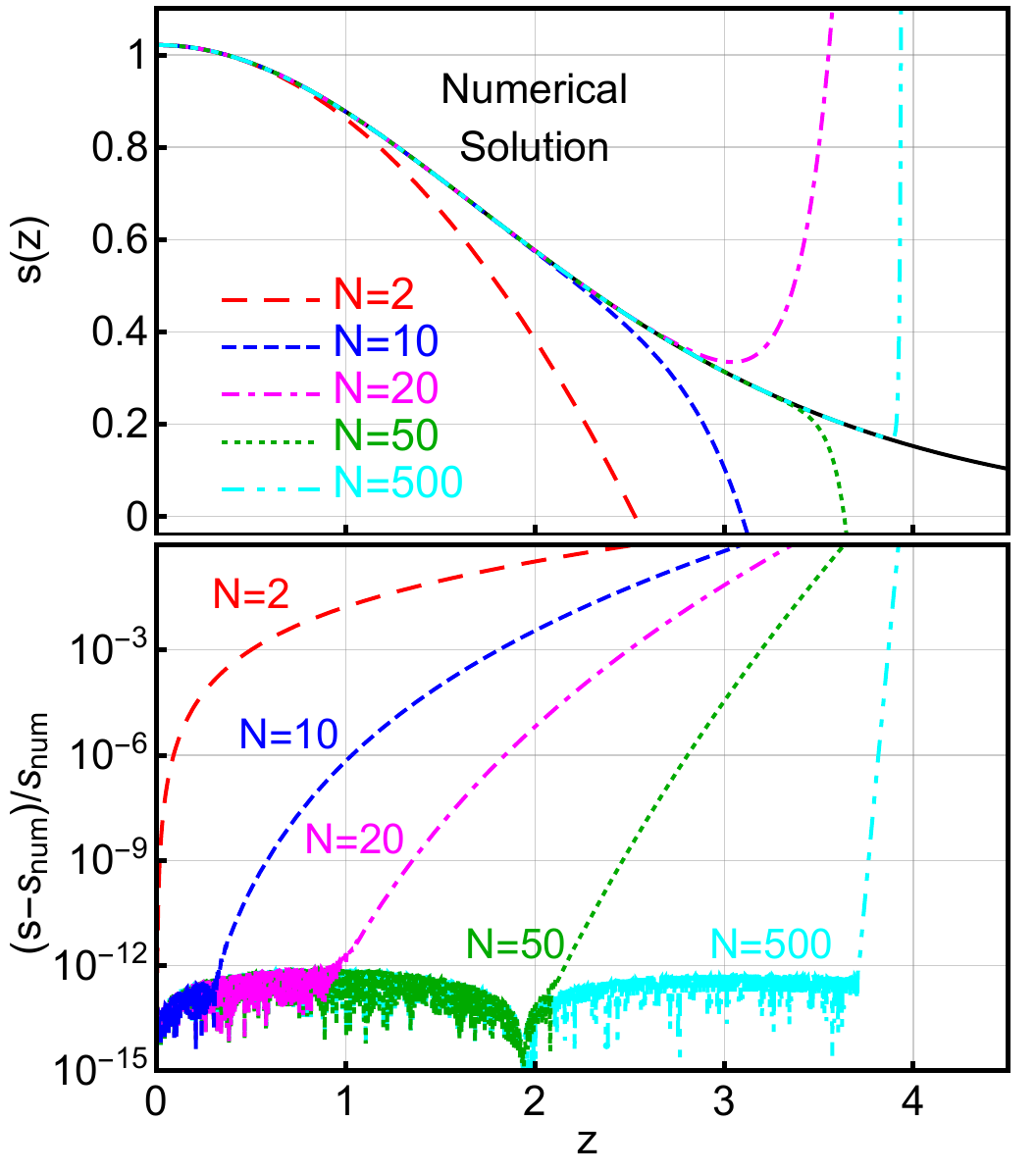}
\caption{The upper panel shows the numerical solution (solid black) and truncated series 
expansion (colored dashed lines) of the central wavefunction. The lower panel shows the 
accuracy $(s_{(N)}-s_{num})/s_{num}$ of the truncated series expansion with respect to the 
numerical solution.}
\label{smallz-s}
\end{figure}

For practical purposes, we will truncate the infinite series of Eq. \ref{smallz-expansion} at 
some $N$. This is shown in Fig. \ref{smallz-s}. The upper panel shows both the numerical solution 
as well as the truncated expansion $s_{(N)}$ for different values of $N$. The lower panel shows 
the deviations of the truncated expansion from the numerical solution. We can see that already a 
small number of terms in the series expansion provides a sub-percent level accuracy for the inner part 
of the profile. A better accuracy can be obtained by including more terms in the expansion. 
However, the accuracy of the series expansion is limited by the accuracy of $s_0$ and $v_0$ 
which in Fig. \ref{smallz-s} is about $10^{-12}$. Note that the series expansion of 
Eq. \ref{smallz-expansion} diverges at $z>4$ and a different parametrization has to be chosen. 

\subsection{Expansion at Large Radius: $z=\infty$}

At large radius $z \to \infty$, we expect the wavefunction to decrease at least exponentially, 
$\psi \sim e^{-kr}$, and the potential to approach the Newtonian potential of a central point 
mass $\Phi \approx \frac{GM}{r}$. A series expansion at large radius must be able to reproduce 
these limits. Let us choose the following general ansatz for the form of the solution
\be
s =\hspace{-0.3cm} \sum_{n,m=0,0}^{\infty,\infty}\hspace{-0.3cm} s^n_m 
\left( \frac{e^{-z}}{z^\sigma} \right)^n \hspace{-0.2cm} z^{-m}  , \; \; \; 
v =\hspace{-0.3cm} \sum_{n,m=0,0}^{\infty,\infty}\hspace{-0.3cm} v^n_m 
\left( \frac{e^{-z}}{z^\sigma} \right)^n \hspace{-0.2cm} z^{-m} 
\label{largez-expansion}
\ee
Here we assume the existence of a $\sigma \in \mathbb{R}$, whose meaning will 
become clear later. The Laplacian of $s$ can be written as
\be
&\hspace{-0.2cm}\nabla^2 s
= \hspace{-0.3cm} \sum_{n,m=0,0}^{\infty,\infty} 
\Big(s^n_m  n^2  
+ 2 s^n_{m-1} n (n\sigma+m-2) \\
&+ s^n_{m-2} (n\sigma+m-3)(n\sigma+m-2)  \Big)  \hspace{-0.15cm} 
\left(\frac{ e^{-z}}{z^\sigma}\right)^n \hspace{-0.2cm} z^{-m} 
\label{largez-laplacian}
\ee
Note that we introduced the short-hand notation $s_{n,-1}=s_{n,-2}=0$.   
The right hand side of Eq.~\ref{gpp-scaleinvariantgpp}  can be rewritten as Cauchy product:
\be
sv 
&=
 \left[\sum_{i,a=0,0}^{\infty,\infty} \hspace{-0.2cm}s^i_a 
\left(\frac{e^{-z}}{z^\sigma} \right)^i \hspace{-0.15cm}z^{-a} \right] \hspace{-0.1cm}
 \left[\sum_{j,b=0,0}^{\infty,\infty} \hspace{-0.2cm}v^j_b \left(\frac{e^{-z}}{z^\sigma} \right)^j 
\hspace{-0.15cm}z^{-b} \right]\\
&= \sum_{n,m=0,0}^{\infty,\infty} 
\left( \sum_{p,q=0,0}^{n,m} s^p_q\;  v^{n-p}_{m-q}\right) \left(\frac{e^{-z}}{z^\sigma}\right)^n z^{-m}
\ee
By matching the coefficients, Eq.~\ref{gpp-scaleinvariantgpp} we obtain the recursion relations
\begin{eqnarray}
\nonumber \hspace{-0.4cm} \sum_{p,q=0,0}^{n,m} &s^p_q v^{n-p}_{m-q} + n^2 \, s^n_m  
+ 2 n(n\sigma+m-2) \; s^n_{m-1}  \\
&+  (n\sigma+m-2)(n\sigma+m-3)s^n_{m-2} =0
\label{largez-eqs}\\
\nonumber \hspace{-0.4cm} \sum_{p,q=0,0}^{n,m} &s^p_q s^{n-p}_{m-q} + n^2 \, v^n_m  
+ 2 n(n\sigma+m-2) \; v^n_{m-1}  \\
&+ (n\sigma+m-2)(n\sigma+m-3)v^n_{m-2} 
= 0. 
\label{largez-eqv}
\end{eqnarray}
Let us note the following properties of $s$ and $v$: i) Normalizability requires $s^0_0=0$. 
Eq.~\ref{largez-eqs} then implies that all coefficients $s^0_m$ vanish as well. This means 
that the wavefunction decays at least exponentially. ii) Eq.~\ref{largez-eqv} then implies 
that all $v^0_m=0$ for $m>1$. This means that at large radius, the potential is described by 
the Newtonian potential $v^{(0)}= v^0_0+\frac{v^0_1}{z}$. All other terms in the expansion of 
$v$ are at least exponentially suppressed. iii) Eq.~\ref{largez-eqs} and \ref{largez-eqv} further 
imply that the potential contains only non-vanishing components $v^n_m$ for even $n$ while the 
wavefunction only has non-vanishing components $s^n_m$ for odd $n$. 

\section{The Solution for the Wavefunction}
\subsection{The Wavefunction at Leading Order}

Using Eq.~\ref{largez-eqs} and \ref{largez-eqv}, we are able to recursively calculate all 
coefficients in the expansion of $s$ and $v$. Let us first analyze the $n=1$ expansion of 
$s$ which provides us both a leading order approximation and a deeper understanding about 
the form of the solution. We have seen before that at leading order the potential is given 
by $v^{(0)}=v^0_0+\frac{v^0_1}{z}$. Setting $m=0$, Eq.~\ref{largez-eqs} reads 
$ s^1_0  = - s^1_0 v^0_0 $ which implies $v^0_0 = -1$ as expected from our scale choice 
which fixes $v(\infty)=-1$. Setting $m=1$, Eq.~\ref{largez-eqs} reads 
\be
s^1_1 + 2 (\sigma-1) s^1_0   = - s^1_1 v^0_0  - s^1_0 v^0_1 
\ee
which implies $v^0_1 = 2 (1-\sigma)$. This is a remarkable result: the exponent $\sigma$ in the 
series expansion is related to the the total mass of the system. In the notation of 
Eq. \ref{gpp-rescale}, we can write $\sigma =1-\frac{1}{2} v^0_1 = 1-\beta $. Let us 
now calculate the remaining coefficients by setting  $m=M+1$ with $M\ge 1$. Then 
Eq. \ref{largez-eqs} can be written as
\be
s^1_{M+1}& v^0_0 + s^1_M v^0_1 + s^1_{M+1} + 2(\sigma+M-1)  s^1_M\\
&  +  (\sigma+M-1)(\sigma+M-2)s^1_{M-1} = 0  
\ee
We can therefore recursively compute the coefficients $s^1_M$ by
\be
s^1_M  = - \frac{ (\sigma + M -1) (\sigma + M -2)}{2M} s^1_{M-1} . 
\ee
Using the rising factorials $(x)_n = x (x+1) (x+2) \cdots (x+n-1)$, $(x)_0=1$ we can write the 
coefficients explicitly as
\be
s^1_M = s^1_0 \frac{(\sigma)_M (\sigma-1)_M}{M!} \cdot (-2)^{-M} 
\label{leadingorder-coeff}
\ee
The leading order wavefunction $s^{(1)}$ can therefore be written as 
\be
s^{(1)} = \alpha \; \frac{e^{-z} }{z^{1-\beta}} \sum_{m=0}^{\infty} 
\left(  \frac{(1-\beta)_m (-\beta)_m}{m!} \; (-2z)^{-m} \right)
\label{leadingorder-expansions}
\ee
Here we have introduced the normalization parameter $\alpha=s^1_0$. The 
far-field solution is described by only two free parameters: the wavefunction 
normalization $\alpha$ and the total mass parameter $\beta$. At very large radius, 
the wavefunction approaches $s = \alpha e^{-z} z^{\beta-1}$.

Let us note that we can rewrite the far-field solution as
\be
s^{(1)}= \frac{ \alpha}{2^{\beta}z} W_{\beta,-\frac{1}{2}}(2z) 
= \frac{\alpha}{ 2^{\beta} z} e^{-z} U(-\beta,0,2z) 
\label{leadingorder-whittaker}
\ee
where $W_{\beta,-\frac{1}{2}}(2z)$ is the Whittaker function which can also be expressed in 
terms of the confluent hypergeometric function $U$. This result is not surprising: when 
considering the leading order potential $v^{(0)}=-1 + \frac{2\beta}{z}$, Eq. \ref{gpp-scaleinvariantgpp} 
turns into the Whittaker equations. For a more detailed discussion on different representations of the 
leading order wavefunction, see appendix \ref{app-whittaker}.

\begin{figure}[t]
\centering
\includegraphics[width=0.45\textwidth]{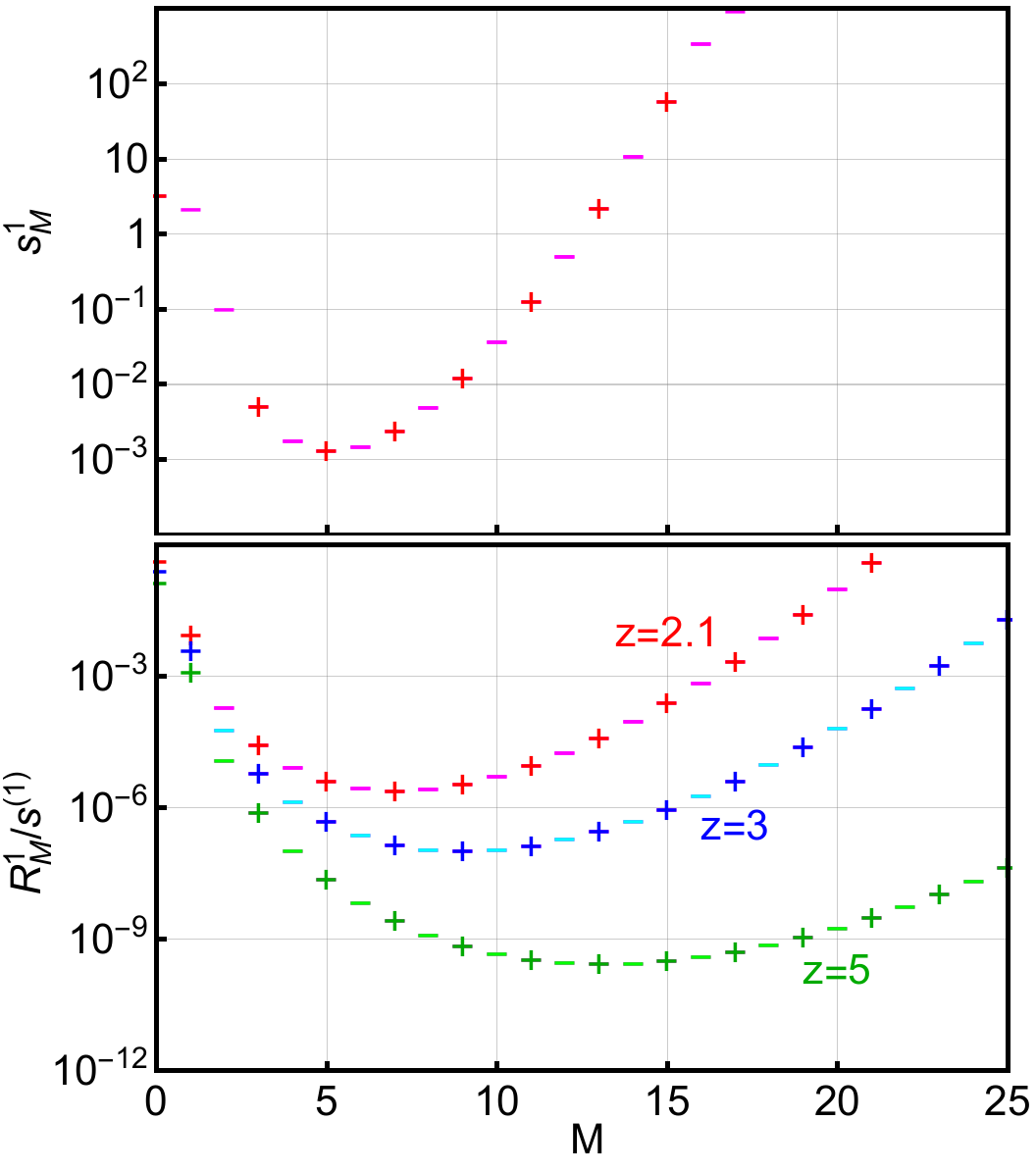}
\caption{Coefficients $s^1_m$ of the leading order series expansion of $s^{(1)}$ as given in 
Eq.~\ref{leadingorder-coeff} (upper panel) and remainder $R^1_M$ as defined in 
Eq.~\ref{leadingorder-remainder} for different values of $z$. In both panels, positive 
values are indicated by a `$+$' while negative values are indicated by a `$-$'. }
\label{leadingorder-coefficient}
\end{figure}

It is  known that the series expansion of the confluent hypergeometric function $U(a,b,z)$, and 
therefore also the expansion of $s^{(1)}$, are not Cauchy convergent. However, the infinite sum has 
a finite value and behaves convergently for a finite number of terms. This is illustrated in the upper 
panel of Fig. \ref{leadingorder-coefficient} where we show the coefficients $s^1_m$. We can see that 
they converge for $m<6$ while they start to diverge again for $m>6$. Let us therefore split the 
function $s^{(1)}$ into a finite series $s^{(1)}_{(M)}$ truncated after $M$ terms and the 
corresponding remainder $R_M$:
\be
s^{(1)} = s^{(1)}_{(M)} + R^1_M \;\;\text{with}\;\; s^{(1)}_{(M)}= 
\sum_{m=0}^{M} s^1_m \frac{e^{-z}}{z^{\beta-1}} z^{-m}  
\label{leadingorder-remainder}
\ee

In the right panel of Fig. \ref{leadingorder-coefficient} we show the relative size of the 
remainder $R^1_M$ with respect to $s^{(1)}$ for different values of $z$. We can see that 
finite series first converges quickly, even for values of $z$ close the convergence 
radius $z>2$. At some $M$, which depends on the value of $z$, the remainder reaches a 
minimum and diverges for large $M$. Note that the coefficients $s^1_m$ and therefore 
also the remainders $R_M$ are oscillating which allows the infinite series $s^{(1)}$ 
to be finite. In this work, we will avoid the problem of divergence by truncating the 
series expansion of $s^{(1)}$ at $m=M$ and ignoring the remainder. This precision of this 
approximation should be sufficient for most applications\footnote{As shown in \cite{Daalhuis1992}, it 
is possible to perform a hyperasymptotic expansions in which we truncate the series at $m=M$ and 
perform another series expansion for the remainder. This procedure can be repeated until the 
desired precision is reached.}. 

Let us now compare the leading order approximation with the numerical solution. To obtain the 
expansion parameters $\alpha$ and $\beta$, we perform a fit of the leading order expansion to 
the numerical solution, as explained in appendix~\ref{app-numerics}. In the upper panel of 
Fig. \ref{leadingorder-truncation} we compare the full leading order approximation $s^{(1)}$ and 
the truncated series $s^{(1)}_{(M)}$ for different $M$ to the numerical solution. We can see 
that the full leading order series expansion $s^{(1)}$ converges to the numerical profile for $z>2$ 
and is already well described by the truncated series with $M=1$. 

The lower panel shows the normalized differences between the truncated series expansion and the 
full leading order solution, $(s^{(1)}_{(M)}-s^{(1)})/s^{(1)}$, and deviation of the full leading 
order solution compared to the numerical solution. For increasing $z$, the differences between the 
numerical solution and the leading order solution $s^{(1)}$ become exponentially small. 
For $z \gtrsim12$ we can see that the precision of the leading order approximation $s^{(1)}$ starts 
to be limited by the precision of the expansion parameters $\alpha$ and $\beta$, which we estimated 
to be of the order $10^{-9}$. The colored dashed lines show the remainder $R^1_M$ of the truncated 
series expansion $s^{(1)}_{(M)}$. We can see the for $M=4$ the uncertainty due to truncation is 
comparable to the uncertainty of the input parameters. 

\begin{figure}[t]
\centering
\includegraphics[width=0.45\textwidth]{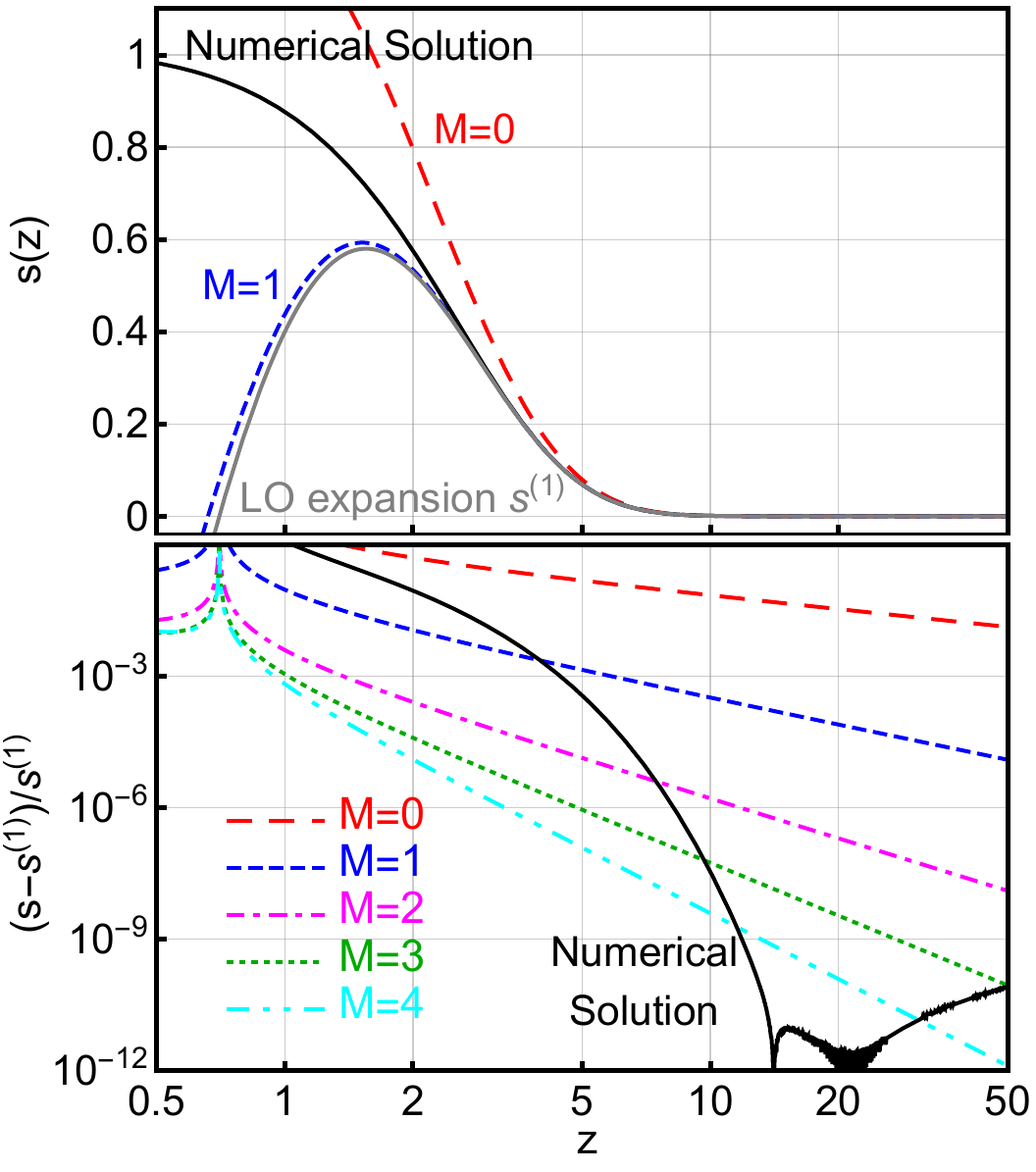}
\caption{The upper panel shows the numerical solution (solid black line), the truncated leading order 
series expansion $s^{(1)}_{(M)}$ (colored dashed lines) and the full leading order series expansion
 $s^{(1)}$ (solid gray line) of the wavefunction. The lower panel shows the accuracy 
$(s^{(1)}_{(M)}-s^{(1)})/s^{(1)}$ of the truncated leading order series expansion with respect 
to the full leading order series expansion, as indicated by the colored dashed lines. The solid 
black line in the lower panel shows the accuracy of the leading order series expansion with respect 
to the numerical solution: $(s_{num}-s^{(1)})/s^{(1)}$.}
\label{leadingorder-truncation}
\end{figure}

At small $z$, we can see that difference between the numerical solution an $s^{(1)}$ increase and 
higher order terms $N>1$ start to be important. 

\subsection{Next to Leading Order Contributions}

\begin{figure}[t]
\centering
\includegraphics[width=0.45\textwidth]{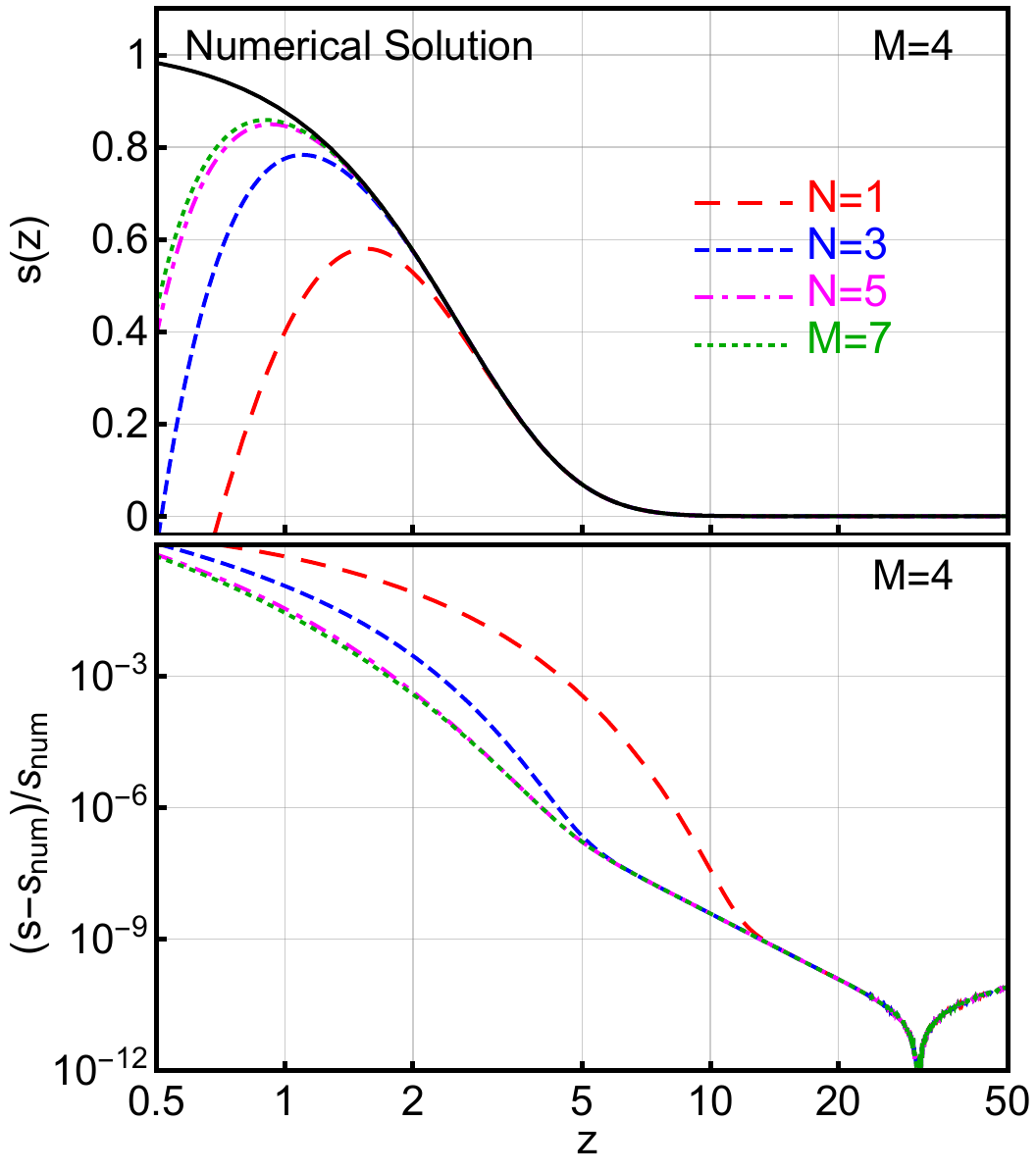}
\caption{The upper panel shows the numerical solution (solid black) and truncated series expansion 
$s^{(N)}_{(M)}$ of the wavefunction. The lower panel shows the accuracy $(s^{(N)}_{(M)}-s_{num})/s_{num}$ 
of the truncated series expansion with respect to the numerical solution. We fix $M=4$.}
\label{nlo-truncation}
\end{figure}

In the previous section we have analyzed the leading order $N=1$ contribution of the series 
expansion in Eq.~\ref{largez-eqs} to the wavefunction. We found that at large $z \gtrsim 12$, the 
contribution from next-to-leading order terms $N>1$ is smaller than the uncertainty induced by the 
uncertainty of the parameter $\alpha$ and $\beta$. We concluded that in this range the $N>1$ terms 
can be safely ignored. However, at intermediate $z$ in the range $ 2 < z < 12$, the next to leading 
order terms become important, as we have seen in Fig.~\ref{leadingorder-truncation} and terms of higher 
order in $N$ have to be included. Let us introduce the truncated solution $s^{(N)}_{(M)}$ and the 
corresponding remainder $R^N_M$ via
\be
\hspace{-0.25cm} s = s^{(N)}_{(M)} + R^N_M \;\;\text{with}\;\; 
s^{(N)}_{(M)}= \hspace{-0.3cm} \sum_{n,m=0,0}^{N,M} \hspace{-0.2cm}s^n_m 
\left(\frac{e^{-z}}{z^{\beta-1}}\right)^n \hspace{-0.15cm} z^{-m}  
\label{nlo-remainder}
\ee
In the upper panel of Fig.~\ref{nlo-truncation} we show the truncated expansion 
$s^{(N)}_{(M)}$ for the wavefunction for different choices of $N$, fixing $M=4$. The lower panel 
shows the corresponding deviations of the truncated expansion from the numerical solution. We can 
see that the series always diverges for $z<1$. Including the $N=3$ and $N=5$ terms will increase 
the accuracy of our series expansion for an intermediate radius $z \approx 3$ to an 
$\mathcal{O}(10^{-5})$ level. Taking into account additional terms $N>5$ does not significantly 
increase the accuracy of the expansion. In this case, the dominant contribution to the remainder 
comes from neglected terms with $m>M$, which become more important at small $z$. 

\subsection{Combined Result and Matching}

\begin{figure}[t]
\centering
\includegraphics[width=0.45\textwidth]{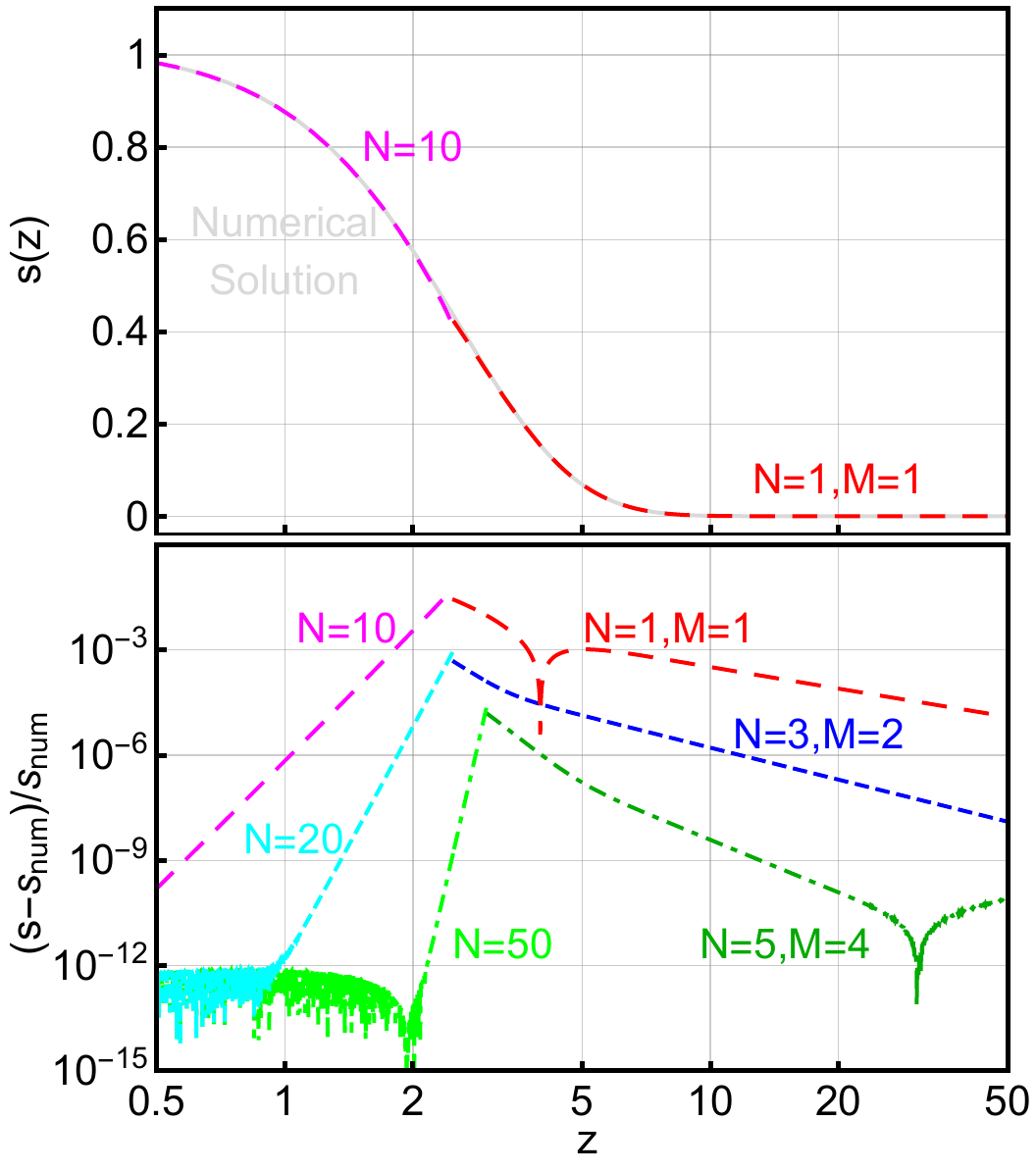}
\caption{The upper panel shows the truncated series expansion of the wavefunction $s_{10}$ at small 
radius (magenta curve) and $s^{1}_{1}$ at large radius (red curve), matched at an intermediate 
value of $z$. The solid gray line shows the numerical solution. The lower panel shows the accuracy 
of the truncated series expansion with respect to the numerical solution for different truncations $N,M$.}
\label{match-solution}
\end{figure}

We can now combine the two truncated solutions, $s_{(N)}$ and $s^{(N)}_{(M)}$, obtained for both small 
and large values of $z$ by matching them at a matching point $z^*$. This is shown for $s_{(N)}$ 
with $N=10$ and $s^{(N)}_{(M)}$ with $N=1$, $M=1$ in the upper panel of Fig.~\ref{match-solution}. Here the 
truncated solution takes the simple form 
\be
\hspace{-0.3cm}
s= \begin{cases}
1.021 -0.159 z^2 + 1.63 \cdot 10^{-2}z^4  & \multirow{3}{*}{$\text{ for } z<2.5$}\\
\;\, -1.42  \cdot 10^{-3} z^6 +1.14  \cdot 10^{-4} z^8 &  \\
\;\, - 8.74 \cdot 10^{-6} z^{10} &  \\ &\\
e^{-z} \cdot z^{0.7526} \left(3.4951 - \frac{2.3053}{z}\right) & \text{ for } z>2.5 \\
\end{cases}
\ee
We can see that already such few terms in the series expansion are sufficient to describe the 
wavefunction well. The accuracy is at the few percent level at the matching point $z=2.5$ and 
orders of magnitude better at small and large $z$, as shown in the lower panel. A precision of 
$10^{-5}$ at the matching point can be achieved using $s_{(N)}$ with $N=50$ and $s^{(N)}_{(M)}$ 
with $N=5$, $M=4$, where the precision at small and large $z$ is again limited by the precision 
of the expansion parameters $s_0$, $v_0$, $\alpha$ and $\beta$ at the $10^{-12}$ level.

We can also determine the expansion parameters by matching the small and large radius wavefunction 
and their derivatives at a matching point $z^*$. We have performed such a matching using $s_{(N)}$ 
with $N=250$ and $s^{(N)}_{(M)}$ with $N=10$ and $M=6$ and obtain
\be
s_0   = 1.0215035 \pm 4.46\cdot10^{-6}\\ 
v_0   = 0.9383304 \pm 2.67\cdot10^{-6}\\ 
\alpha= 3.4951958 \pm 3.17\cdot10^{-5}\\
\beta = 1.7526505 \pm 6.34\cdot10^{-6}
\ee
To estimate the uncertainty associated with the matching procedure, we performed multiple matchings 
for $3<z^*<3.5$. The obtained values for expansion parameters are consistent with those in 
Eq.~\ref{numerics-paramaters} obtained through the fitting of the large radius solution but have 
a significantly worse precision. This is not surprising, since the precision of our series 
expansion is expected to be worst at the matching point $z^*$. 


\section{Conclusion}

We have found a semi-analytic solution to the coupled Poisson-Newton 
equations describing dilute boson stars. We have shown that our solution is stable to numerical errors, and
that it reproduces the numerical results with accuracy better than $10^{-5}$ over the entire
range. Improvements in accuracy can easily be attained for a small expense of numerical work. 

There are many possible applications of our methods. The simplest one is to consider interacting
bosons, when a potential for the bosons is added. Such a potential can significantly
modify the solution, because the interactions can be much stronger than gravitational. 
Rotation can also modify the solution. In all these case, the large number of parameters makes it impractical to find purely numerical solutions; our semi-analytical method would be better suited for these problems.  

Another open question is the stability of these solutions.
For example, it is not known how the boson stars are affected by external perturbations e.g. by another 
star nearby. Once again, an accurate knowledge of the profiles is a requirement for
the stability analysis. We hope to return to these and other questions in future work.


\acknowledgments
This work is supported by NSF under Grant PHY-1620638.

\appendix

\section{Numerical Integration and Fitting}
\label{app-numerics}

To obtain a numerical solution, it is convenient to use the variables $S$, $V$ and coordinate $x$ 
given in Eq.~\ref{gpp-rescale} with a reference scale $k$ chosen such that $V(0)=1$. As shown in 
\cite{Tod1992}, the solutions of Eq.~\ref{gpp-dimensionless} can then be parametrized by the central 
value of the wavefunction, $S_0=S(0)$ and categorized into three distinct classes: for $S_0>S_0^*$ 
the wavefunction diverges for at large radius towards positive infinity, for $S_0=S_0^*$ the 
wavefunction converges to zero, is positive definite and square integrable, while for $S_0<S_0^*$ 
the wavefunction diverges for at large radius towards negative infinity. 

Using a Runge-Kutta 4 method with constant step size $\Delta_x$, we perform the numerical integration 
until the wavefunction starts to diverge and iteratively optimize the central value of the wavefunction 
$S_0$ to find $S_0^*$. The precision of the wavefunction needed for the numerical solution to stay 
finite until a large value of $x$, which is needed to fit the large range solution, increases 
exponentially with the radial coordinate $x$. In this study we use a precision of 35 significant 
figures for $S_0$, providing a converging numerical solution for $x < 35$. The accuracy of the 
numerical solution is limited by the step size $\Delta_x$. In this study, we use 
$\Delta_x = 10^{-3}$, providing an accuracy of the solution of order 
$\mathcal{O}\left(\Delta_x^4 \right) \approx 10^{-12}$. 

To obtain the expansion parameters $\alpha,\beta$, we fit the Newtonian potential $V(x) = -k^2 + 
\frac{2k\beta}{x}$ and the Whittaker solution $S(x) =\frac{ k \alpha}{2^\beta x} 
W_{\beta,-\frac{1}{2}}(2kx)$ to the numerical solution for $V$ and $S$ at large $x$. To avoid 
systematic effects due to the truncation of subleading terms $n>1$ of the series expansion in 
Eq.~\ref{largez-expansion}, we restrict the fitting range to $x>x^*$, where the fraction of mass 
outside radius $x^*$ contributed less than $10^{-12}$ to the total mass of the boson star. The 
expansion parameters at small radius can then be obtained through $s_0=k^{-2} S_0$ and $v_0 = k^{-2}$. 
We find that 
\be
s_0    &=  1.02149303631 \pm 1.4 \cdot 10^{-10}\\
v_0    &=  0.93832284019 \pm 1.3 \cdot 10^{-10}\\
\alpha&=  3.4951309897\phantom{9}  \pm 5.1 \cdot 10^{-9}\\
\beta  &=  1.7526648513\phantom{9}  \pm 1.3 \cdot 10^{-9}
\label{numerics-paramaters}
\ee
The uncertainty of $s_0,v_0$ and $\beta$ were estimated by the difference of the best fit values to 
potential and wavefunction. For $\alpha$ we state the uncertainty of wavefunction fit. 

\section{Representations of the Leading Order Wavefunction}
\label{app-whittaker}

When considering only the potential $v^{(0)}=-1 + \frac{2\beta}{z}$, Eq.~\ref{gpp-scaleinvariantgpp} can 
be written as
\be
\frac{d^2}{dz^2}(2z s^{(1)}) + 2z s^{(1)} \left(-1 + \frac{2\beta}{z}\right) =0
\ee
or
\be 
\frac{d^2 w}{dy^2} + \left(-\frac{1}{4} + \frac{\beta}{y} \right) w =0.
\ee
Here we performed a change of variables to $y=2z$ and $w=2z \, s^{(1)}$. This is the Whittaker 
equation \cite{Abramowitz:1974}, which has the solution\footnote{Note that in principle there are 
three additional solutions: (1) $w \sim M_{\beta,\frac{1}{2}}(z)$  which diverges at 
$z \to \infty$, (2) $w \sim M_{\beta,-\frac{1}{2}}(z)$ which is undefined for every value of 
$\beta,z$ and (3) $w \sim W_{\beta,\frac{1}{2}}(z) = W_{\beta,-\frac{1}{2}}(z)$.}  
$w= C \cdot W_{\beta,-\frac{1}{2}}(y)$ and therefore implies $s^{(1)} = C\cdot (2z)^{-1} 
W_{\beta,-\frac{1}{2}}(2z)$. Note that the Whittaker function is related to the confluent 
hypergeometric function $U$ through $W_{\kappa,\mu}(z) = e^{-\frac{1}{2}z} z^{\frac{1}{2}+\mu} 
U(\frac{1}{2}+\mu-\kappa,1+2\mu,z)$ which was used in Eq.~\ref{leadingorder-whittaker}. We can 
expand the Whittaker function in an infinite series
\be
\hspace{-0.2cm} s^{(1)} 
=  \frac{C}{ 2^{1-\beta}} \frac{e^{-z}}{ z^{1-\beta}} \sum_{m=0}^{\infty} 
\left(  \frac{(1-\beta)_m (-\beta)_m}{m!}  (-2z)^{-m} \right)
\ee
This is the same solution as Eq.~\ref{leadingorder-expansions} and we can identify 
$\alpha= C 2^{\beta-1}$. 

The far-field wavefunction can also be written in terms of the Bateman function $k_{\nu}(z)$
\be
\hspace{-0.2cm} 
s^{(1)}=\frac{\alpha}{2^\beta z} e^{-z} U(-\beta,0,2z) = \frac{\alpha}{2^\beta z} \Gamma(1+\beta) k_{2\beta}(z).
\ee
The Bateman function \cite{Batemann1931} can be expressed via the integral form
\be
k_{2\beta}(z) = \frac{2}{\pi} \int_0^{\frac{\pi}{2}} \cos\left(z \tan\theta -2\beta\theta\right)d\theta
\ee
which is finite and convergent.


\bibliographystyle{utphys}
\bibliography{references}

\end{document}